\documentclass[structabstract]{aa}
\usepackage{graphicx}
\usepackage{txfonts}
\usepackage{longtable}

\begin{document}
   \title{Detection of an exoplanet around the evolved K giant HD 66141}
   \author{B.-C. Lee\inst{1},
          D.E. Mkrtichian\inst{2,3},
          I. Han\inst{1},
          M.-G. Park\inst{4},
          \and
          K.-M. Kim \inst{1}
          }

   \institute{Korea Astronomy and Space Science Institute, 776,
		Daedeokdae-Ro, Youseong-Gu, Daejeon 305-348, Korea\\
	      \email{[bclee;iwhan;kmkim]@kasi.re.kr}
        \and
        National Astronomical Research Institute of Thailand, Chiang Mai 50200, Thailand
        \and
        Crimean Astrophysical Observatory, Nauchny, Crimea, 98409, Ukraine\\
          \email{davidmkrt@gmail.com}
	    \and
	    Department of Astronomy and Atmospheric Sciences, Kyungpook National University, Daegu 702-701, Korea\\
	      \email{mgp@knu.ac.kr}
             }

   \date{Received 5 September 2011 / Accepted xx xxx 2012}


  \abstract
   {}
   {We have been carrying out a precise radial velocity (RV) survey for K giants to search for and study the origin of the low-amplitude and long-periodic RV variations.
   }
   {We present high-resolution RV measurements of the K2 giant HD 66141 from December 2003 to January 2011 using the fiber-fed Bohyunsan Observatory Echelle Spectrograph (BOES) at Bohyunsan Optical Astronomy Observatory (BOAO).
   }
   {We find that the RV measurements for HD 66141 exhibit a periodic variation of 480.5 $\pm$ 0.5 days with a semi-amplitude of 146.2 $\pm$ 2.7 m s$^{-1}$. The Hipparcos photometry and bisector velocity span (BVS) do not show any obvious correlations with RV variations. We find indeed 706.4 $\pm$ 35.0 day variations in equivalent width (EW) measurements of H$_{\alpha}$ line and 703.0 $\pm$ 39.4 day variations in a space-born measurements 1.25$\mu$ flux of HD 66141 measured during COBE/DIRBE experiment. We reveal that a mean value of long-period variations is about 705 $\pm$ 53 days and the origin is a rotation period of the star and variability that is caused by surface inhomogeneities. For the 480 day periods of RV variations an orbital motion is the most likely explanation. Assuming a stellar mass of 1.1 $\pm$ 0.1 $\it M_{\odot}$ for HD 66141, we obtain a minimum mass for the planetary companion of 6.0 $\pm$ 0.3 $M_\mathrm{Jup}$ with an orbital semi-major axis of 1.2 $\pm$ 0.1 AU and an eccentricity of 0.07 $\pm$ 0.03.
   }
   {}


   \authorrunning{B.-C. Lee et al.}
   \titlerunning{Detection of exoplanet around the evolved K giant HD 66141}
   \maketitle
%

\section{Introduction}

So far, more than 570 exoplanets have been detected and roughly 80\% of them have been found via high resolution spectroscopic measurement of radial velocity (RV) of spectral lines. Besides RV measurements, high resolution spectroscopy simultaneously offers various information on the chemical composition of hosting stars. Among the planetary companions discovered by the RV method, only $\sim$ 30 exoplanets have been detected around giant stars. Cool evolved stars such as K giants are suitable candidates for precise RV measurements because they have sharp and sufficient spectral lines compared to those of early spectral class intermediate-mass and rapidly-rotating main-sequence (MS) stars. As the star moves towards giants with convective outer envelopes, the role of surface inhomogeneities (chromospheric activities or stellar spot modulations) increase. Thus, it is difficult to distinguish the RV variation due to a planetary companion from that due to an inhomogeneity in evolved stars and therefore, giants still remain an undeveloped region in exoplanet surveys.
The suggestion of planetary companions in K giants (Hatzes \& Cochran 1993) was first confirmed by Hatzes et al. (2005, 2006) and Refferet et al. (2006) after 13 years of observation.

Currently, a number of groups are conducting exoplanet surveys around giants (Frink et al. 2002; Sato et al. 2003; Setiawan et al. 2003; Hatzes et al. 2005; Johnson et al. 2007; Lovis \& Mayor 2007; Niedzielski et al. 2007; D{\"o}llinger et al. 2009; Han et al. 2010).
For the past eight years, we have conducted precise RV measurements of 55 K giants. Here, we present a long-period and low-amplitude RV variation of a K giant HD 66141. Data observations and analysis are presented in Sect. 2. In Sect. 3 and Sect. 4, we describe the properties of HD 66141 and analyze the period search, respectively. The nature of the RV variations is investigated in Sect. 5. Finally, we discuss the work presented in this paper in Sect. 6.

%

\section{Observations and analysis}

We started a precise RV survey using 1.8-m telescope at BOAO in 2003 to search for exoplanets and to study the asteroseismology of 55 K giants. We have reported two new exoplanets (Han et al. 2010; Lee et al. 2011), a confirmation of an exoplanet (Han et al. 2008), and an oscillating star (Kim et al. 2006).

We acquired 54 spectra for HD 66141 (= HR 3145 = HIP 39311) from December 2003 to January 2011 using the fiber-fed high-resolution ($\emph{R}$ = 90 000) Bohyunsan Observatory Echelle Spectrograph (BOES; Kim et al. 2007) attached to the 1.8-m telescope at Bohyunsan Optical Astronomy Observatory (BOAO) in Korea. An iodine absorption cell (I$_{2}$) was used to provide the precise RV measurements. Each estimated signal-to-noise ratio (S/N) at the I$_{2}$ wavelength region is about 200 -- 250, with typical exposure time ranging between 240 and 480 seconds.
The RV measurements for HD 66141 are listed in Table~\ref{tab2}.

The reduction was carried out using IRAF (Tody 1986) software and DECH codes (Galazutdinov 1992). IRAF was used for image processing and the extraction of 1--D spectra, and a continuum process was performed using the DECH code. I$_2$ analyses and precise RV measurements were undertaken using the RVI2CELL (Han et al. 2007) developed at the Korea Astronomy $\&$ Space Science Institute (KASI). The RVI2CELL adopts basically the same algorithm and procedures described by Butler et al. (1996). However, for modeling the instrument profile we used the matrix formula described by Endl et al. (2000). We solved the matrix equation using singular value decomposition instead of the maximum entropy method adopted by Endl et al. (2000). The code have been used in several cases (Kim et al. 2006; Han et al. 2008, 2010; Lee et al. 2008, 2011)

The RV standard star $\tau$ Ceti shows an rms scatter of 6.7 m s$^{-1}$ over the time span of our observations and demonstrates the long-term stability of the BOES (Lee et al. 2011).

%

\section{The properties of HD 66141}

%
\begin{table}
\begin{center}
\caption[]{Stellar parameters for HD 66141.}
\label{tab1}
\begin{tabular}{lcc}
\hline
\hline
    Parameter          & Value      &    Reference     \\

\hline
    Spectral type            & K2 III    & Perryman et  al.(1997)  \\
    $\textit{$m_{v}$}$ (mag) & 4.39      & Perryman et  al.(1997)  \\
    $\textit{$M_{v}$}$ (mag) & -- 0.15   & Perryman et  al.(1997)  \\
    $\textit{B-V}$ (mag)     & 1.26      & Massarotti (2008) \\
    age (Gyr)                & 6.84 $\pm$ 1.39\tablefootmark{a} & Derived  \\
    Distance (pc)            & 80.9 $\pm$ 6.3   & Famaey et al. (2005) \\
                             & 80               & Massarotti et al. (2008) \\
    RV (km s$^{-1}$)         & 71.57 $\pm$ 0.01 & Famaey et al.  (2005) \\
    Parallax (mas)           & 12.49 $\pm$ 0.98  & Perryman et  al.(1997)   \\
                             & 12.48             & Kharchenko et al. (2007) \\
                             & 12.84 $\pm$ 0.25  & van Leewen (2007)  \\
    Diameter (mas)           & 2.81 $\pm$ 0.15   & Richichi \& Percheron (2002) \\
    $T_{\rm eff}$ (K)        & 4305              & Massarotti  et al. (2008) \\
                             & 4323  $\pm$ 15    & This work                 \\
    $\rm [Fe/H]$             & -- 0.36           & Massarotti  et al. (2008) \\
                             & -- 0.323 $\pm$ 0.034  & This work      \\
    log $\it g$             & 2.3                & Massarotti et al. (2008) \\
                            & 1.92 $\pm$ 0.07    & This work \\
                            & 1.78 $\pm$ 0.04\tablefootmark{a}    & Derived \\
    $\textit{$R_{\star}$}$ ($R_{\odot}$) & 22               & Massarotti  et al. (2008) \\
                                         & 21.4 $\pm$ 0.6\tablefootmark{a}   & Derived \\
    $\textit{$M_{\star}$}$ ($M_{\odot}$) & 1.1  $\pm$ 0.1\tablefootmark{a}   & Derived \\
    $\textit{$L_{\star}$}$ ($L_{\odot}$) & 174   & Massarotti  et al.  (2008)  \\
    $v_{\rm rot}$ sin $i$ (km s$^{-1}$)  & 2.5            & Fekel (1997)                \\
                                         & 1.1 $\pm$ 1.0  & de Medeiros \& Mayor (1999) \\
                                         & 4.7            & Massarotti  et al. (2008)   \\
    $P_{\rm rot}$ / sin $i$ (days)      & 253 -- 938  & This work   \\
    $v_{\rm micro}$ (km s$^{-1}$)       & 1.45  $\pm$ 0.08  &  This work   \\

\hline

\end{tabular}
\end{center}
\tablefoottext{a}{Derived using an online tool (http://stevoapd.inaf.it/cgi-bin/param}).
\end{table}
HD 66141 is an IAU bright (V = 4.39, K2 III) RV standard star adopted by IAU Commission 30 (Pearce 1955).
It has been observed very frequently since and multiple RVs of the object are available in the literature (e.g.
RV = 71.6 $\pm$ 0.3 m s$^{-1}$ [Udry et al. 1999a]; 71.40 $\pm$ 0.15 m s$^{-1}$ [Eaton \& Williamson 2007]). The RV stability of HD 66141 and other IAU \emph{standards} were indeed under doubts during last decades (Batten 1983). Over 20 years of its and other IAU standard observation with CORAVEL were summarized in Udry et al. (1999a). While, authors did not find a regular variability of RVs in HD 66141, Udry et al. (1999b) did not include it to a list of recommended standards for future use due to the fact that as a giant it exhibits some RV variability at higher precision level.

We determined the atmospheric parameters of HD 66141 directly from our spectra. By using 266 measured equivalent widths (EWs) of Fe I and Fe II lines, we determined $T_{\mathrm{eff}}$, [Fe/H], log $\it g$, and
$v_{\mathrm{micro}}$ of the star using the program TGVIT (Takeda et al. 2005). To estimate the stellar radius, mass and surface gravity, we used an online tool (http://stev.oapd.inaf.it/cgi-bin/param), which is based on theoretical isochrones (Girardi et al. 2000; J{\o}rgensen \& Lindegren 2005; da Silva et al. 2006). The result is $\textit{$R_{\star}$}$ = 21.4 $\pm$ 0.6 $R_{\odot}$, log $\it g$ = 1.78 $\pm$ 0.04, and $\textit{$M_{\star}$}$ = 1.1 $\pm$ 0.1 $M_{\odot}$.

The $v_{\mathrm{rot}}$ sin $i$ values of HD\,66141 are from  de Medeiros \& Mayor(1999),  Massarotti et al. (2008) and Fekel (1977). Two former author measured the rotational velocities using the spectroscopic line broadening methods and the cross-correlation spectrometers. Fekel (1977) used a high-resolution spectra. These determinations were also corrected for macroturbulence and instrumental broadening.
The rotational velocity of 2.5 km s$^{-1}$ (Fekel 1997) is close to the 1.1 $\pm$ 1.0 km s$^{-1}$ (Medeiros \& Mayor 1999) but different with determination of 4.7 km s$^{-1}$ (Massarotti et al. 2008).
With these determinations we have a range of uncertainty of $v_{\mathrm{rot}}$ sin $i$ measurements 1.1 -- 4.7 km s$^{-1}$. Adopting a stellar radius of 21.44 $\pm$ 0.58 $R_{\odot}$ and ignoring the sin $i$ we get the
uncertainty range for the upper limit of the rotation period

   $P_{\mathrm{rot}} = 2 \pi R_{\star}$ / ($v_{\mathrm{rot}}$ sin $i$) = 253 -- 938 days.\\

\hskip -15pt
The basic stellar parameters for HD 66141 are summarized in Table~\ref{tab1}.

%

\section{Period search}

\begin{table}
\begin{center}
\caption{RV measurements for HD 66141 between December 2003 and January 2011.}
\label{tab2}
\begin{tabular}{cccccc}
\hline\hline

 JD         & $\Delta$RV  & $\pm \sigma$ &        JD & $\Delta$RV  & $\pm \sigma$  \\
 -2 450 000 & m\,s$^{-1}$ &  m\,s$^{-1}$ & -2 450 000  & m\,s$^{-1}$ &  m\,s$^{-1}$  \\
\hline

2976.251556  &    -49.8   &     9.7  &   3459.015331  &    -59.9   &     7.3   \\
2977.315941  &   -123.3   &     8.1  &   3459.080221  &    -74.1   &     7.7   \\
2977.324622  &   -123.6   &     8.1  &   3729.107766  &    189.1   &     9.2   \\
2978.186977  &    -35.9   &     9.1  &   3759.164025  &    132.2   &     7.2   \\
2978.205694  &    -36.5   &     9.0  &   3778.200411  &    142.8   &     8.2   \\
2980.409693  &    -75.4   &    14.6  &   3779.170158  &    167.8   &     7.4   \\
3045.039195  &    -15.0   &     6.7  &   3818.009412  &     40.4   &     6.7   \\
3045.047343  &    -16.6   &     6.7  &   3819.971947  &     64.5   &     7.7   \\
3046.104835  &    -51.8   &     9.0  &   4036.337614  &     31.3   &     6.5   \\
3046.115923  &    -48.9   &     7.9  &   4123.092626  &    162.7   &     7.2   \\
3046.123145  &    -45.7   &    11.2  &   4126.108709  &    172.3   &     6.5   \\
3048.085051  &      0.8   &     7.3  &   4214.007594  &    203.0   &    10.8   \\
3048.092203  &      8.9   &     8.3  &   4396.340192  &   -125.4   &     6.7   \\
3072.070064  &      5.0   &    10.2  &   4452.360964  &   -146.5   &     7.3   \\
3072.078211  &     16.6   &    10.9  &   4506.168231  &    -79.0   &     7.2   \\
3095.998435  &     16.9   &     7.1  &   4536.065764  &     -4.7   &     6.7   \\
3132.993871  &     76.4   &     7.2  &   4538.042117  &     18.3   &     6.5   \\
3332.318368  &    -33.1   &     7.3  &   4755.368868  &     27.4   &     7.4   \\
3332.325255  &    -35.6   &     6.9  &   4847.220274  &   -134.2   &     6.9   \\
3354.234041  &     34.0   &     6.9  &   4880.152070  &   -144.2   &     8.5   \\
3395.246936  &    -53.8   &    11.9  &   4929.020161  &   -142.7   &     7.1   \\
3430.046636  &    -21.6   &     7.3  &   5171.229126  &    174.3   &     8.2   \\
3432.981185  &    -40.8   &     9.2  &   5248.098927  &     27.2   &     6.4   \\
3432.991011  &    -44.8   &     8.1  &   5251.084871  &     16.6   &     6.6   \\
3433.115135  &    -63.5   &     8.0  &   5456.347802  &     -4.0   &     8.4   \\
3433.126349  &    -58.0   &     7.6  &   5554.319097  &    181.7   &    15.1   \\
3433.183059  &    -68.1   &     8.8  &   5581.132641  &    144.7   &     8.6   \\
\hline

\end{tabular}
\end{center}
\end {table}

%
   \begin{figure}
   \centering
   \includegraphics[width=8cm]{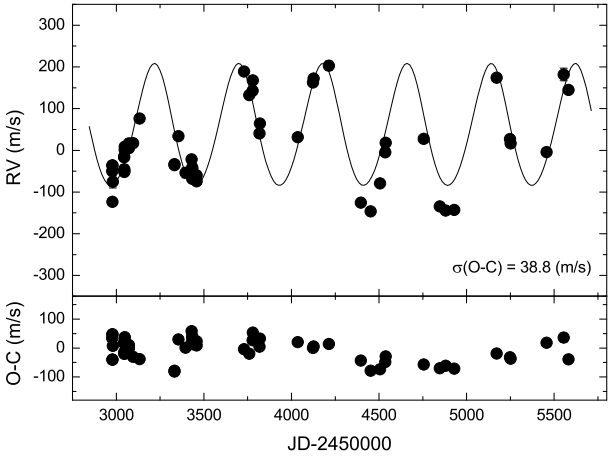}
      \caption{RV curve (\emph{top panel}) and  rms scatter of the residual (\emph{bottom panel}) for HD 66141 from December 2003 to January 2011. The solid line is the orbital solution with a period of 480.5 $\pm$ 0.5 days and an eccentricity of 0.07 $\pm$ 0.03.
              }
         \label{orbit}
   \end{figure}
%

%
   \begin{figure}
   \centering
   \includegraphics[width=8cm]{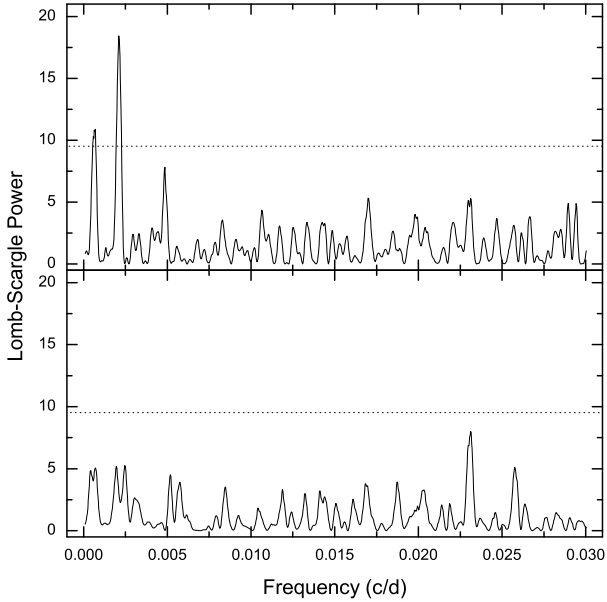}
      \caption{The Lomb-Scargle periodogram of the RV measurements for HD 66141.
      The periodogram shows a significant power at a frequency of 0.002081 $\pm$ 0.000002 c d$^{-1}$
      corresponding to a period of 480.5 $\pm$ 0.5 days (\emph{top panel}) and after subtracting the main frequency variations (\emph{bottom panel}). The horizontal dotted lines indicate an FAP threshold of 1 $\times 10^{-3}$ (0.1\%).
              }
         \label{power1}
   \end{figure}

RV measurements for HD 66141 are presented in Figure~\ref{orbit} and have a standard deviation of 94.5 m s$^{-1}$, which is 14 times larger than the RV standard star $\tau$ Ceti. The long-term periodic variability in the data is indeed visible by eye. The Lomb-Scargle periodogram for unequally spaced data (Lomb 1976; Scargle 1982) was applied to the RV time series of HD 66141 in order to find an accurate period. The frequency uncertainties are standard least-square uncertainties of a sine-wave fit to the data. The periodogram shown in Figure~\ref{power1} (top panel) denotes a significant power at $f_{1}^{\rm RV}$ = 0.002081 $\pm$ 0.000002 c\,d$^{-1}$ ($P_{1}^{\rm RV}$ = 480.5 $\pm$ 0.5 days). We determined the significance of the period by calculating the false alarm probability (FAP) for the dominant period by a bootstrap randomization technique (K\"{u}rster et al. 1999). We computed that among 200 000 trials the highest peak of periodograms, thus we found an FAP of less than $10^{-6}$ for $f_{1}^{\rm RV}$. Our criterion for statistically significant signal in the data is that it must have FAP$<$ $10^{-3}$. We note, that the 480.5 day period is within the uncertainty range 253 -- 938 days for the rotation period found in the previous Section.

Figure~\ref{power1} (bottom panel) shows the Lomb-Scargle periodogram of the residuals after removing the best fit; it exhibits no statistically significant peaks at the domain of interest.


%
   \begin{figure}
   \centering
   \includegraphics[width=8cm]{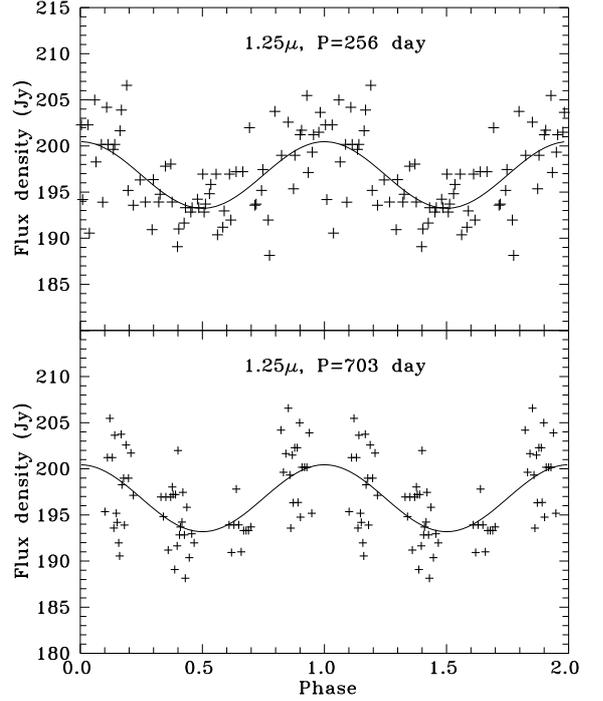}
      \caption{The 1.25$\mu$ flux intensity measurements of HD~66141 phased to the period of 256 (\emph{top panel}) and 703 days (\emph{bottom panel}). Epochs of maximum flux for 256-day and 703-day periods are JD 2447945.3 and JD 2448098.0, respectively.
      }
         \label{cobe}
   \end{figure}

%
%

\section{Origin of the RV variations}

Evolved stars exhibit pulsations as well as surface activity, and they result in low-amplitude RV variabilities on different time scales. While short-term (hours to days) RV variations have been known to be the result of stellar pulsations (Hatzes \& Cochran 1998), long-term (hundreds of days) RV variations with a low-amplitude may be caused by stellar pulsations, rotational modulations by inhomogeneous surface features, or planetary companions.
To establish the origin of the period for HD 66141, we examined
1) the Hipparcos and COBE/DIRBE infrared photometry,
2) the stellar chromospheric activity,
3) the spectral line bisectors,
and 4) the orbital fit.

   \begin{figure}
   \centering
   \includegraphics[width=8cm]{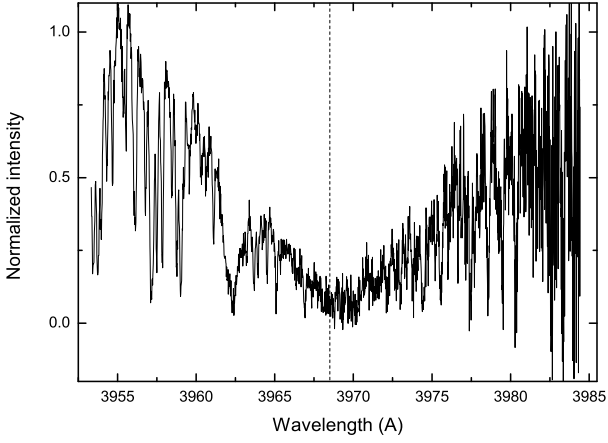}
      \caption{Ca II H spectral region for HD 66141. It shows that the Ca II H core feature exhibits no emission at the line center.
        }
        \label{Ca1}
   \end{figure}
%

%
   \begin{figure}
   \centering
   \includegraphics[width=8cm]{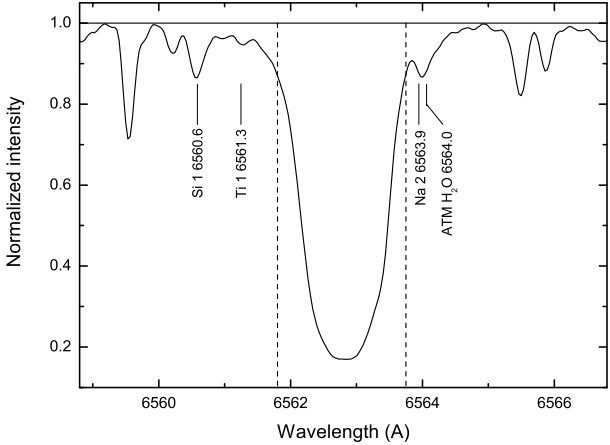}
      \caption{Line profile near H$_{\alpha}$ region for HD 66141. Two vertical dashed lines denote the range of H$_{\alpha}$ to measure the EWs.
        }
        \label{Ha1}
   \end{figure}
\subsection{Hipparcos and COBE/DIRBE infrared photometry}

We analyzed the Hipparcos photometry (ESA 1997) for HD 66141 to search for any possible brightness variations
due to the rotational modulation of stellar spots. For three years, between JD 2447966 and JD 2448754, the Hipparcos satellite obtained 45 photometric measurements for HD 66141 and the data maintained a photometric stability down to the rms scatter of 0.0078 magnitude, corresponding to 0.17\% variations.
Figure~\ref{origin2} shows the Lomb-Scargle periodogram of these measurements. There is no significant peak near the frequency of $f_{1}^{\rm RV}$ = 0.00208 c d$^{-1}$. Two statistically insignificant peaks are visible at frequencies $f_{1}^{\rm HIP}$ = 0.00233 $\pm$ 0.00002 c\,d$^{-1}$ and $f_{1}^{\rm HIP}$ = 0.00274 $\pm$ 0.00002 c d$^{-1}$ with FAPs of $\sim$ 5 $\times 10^{-3}$ that most likely belongs to 1/365.25 day = 0.00274 c\,d$^{-1}$ artifacts. So, we conclude that Hipparcos photometry does not resolve any signals  of surface features like spot.

Nearly at the same time, between JD\,2447973 and JD\,2449128, HD\,66141 was measured in the near-infrared (NIR) 1.25, 2.2, 3.5, and 4.9$\mu$ bands by  NASA's COBE (Cosmic Backgroud Explorer) satellite using DIRBE (Diffuse Infrared Background Experiment) instrument. The total N = 68 weekly averaged fluxes in each band were extracted for HD\,66141 from COBE/DIRBE archives (Price et al. 2010) which we used for analysis. The Lomb-Scargle periodogram analysis of 2.2, 3.5, and  4.9$\mu$ fluxes does not reveal any significant signals in the domain of interest, while the 1.25$\mu$ shows the statistically significant signal (FAP $<$ $10^{-5}$) at the frequency of $f_{\rm 1}^{\rm IR}$ = 0.00390 $\pm$ 0.00008 c\,d$^{-1}$ ($P_{\rm 1}^{\rm IR}$ = 256.1 $\pm$ 5.4 days) with semi-amplitude $\Delta$I = $\pm$ 4.0 Jy. The second statistically significant peak (FAP $<$ $10^{-4}$) is  at the lower frequency of $f_{\rm 2}^{\rm IR}$ = 0.00142 $\pm$ 0.00008 c\,d$^{-1}$ ($P_{\rm 2}^{\rm IR}$ = 703.0 $\pm$ 39.4 days).

After removal from the original data of the 0.00390 or 0.00142 c\,d$^{-1}$ signals the residuals do not show any significant variations (Figure\,\ref{origin2}).
Figure~\ref{cobe} shows the phase curve of the original data folded with two periods $P_{\rm 1}^{\rm IR}$ = 256 days (upper panel) and $P_{\rm 2}^{\rm IR}$ = 703 days (bottom panel). We note that 256-day and 703-day periods are within uncertainty limits for the rotation period of 253 -- 938 days found in Section 3.

\subsection{Chromospheric activity}

The EW variations of Ca II H and H$_{\alpha}$ lines are frequently used as chromospheric activity indicators.
The emissions in the Ca II H core are formed in the chromosphere and show a typical central reversal in the existence of chromospheric activity (Pasquini et al. 1988; Saar \& Donahue 1997).
The existence of extra emission at the center of the line implies that the source function in the chromosphere is larger than in the photosphere. This reversal phenomenon is common in cool stars and is intimately connected to the existence of a convective envelope and magnetic activity.
Unfortunately, the Ca II H line region for HD 66141 does not have enough S/N to estimate EW variations (Figure~\ref{Ca1}). However, it is large enough to check the emission feature in the Ca II H line core, and we did not find any emission.

\begin{figure}
   \centering
   \includegraphics[width=8cm]{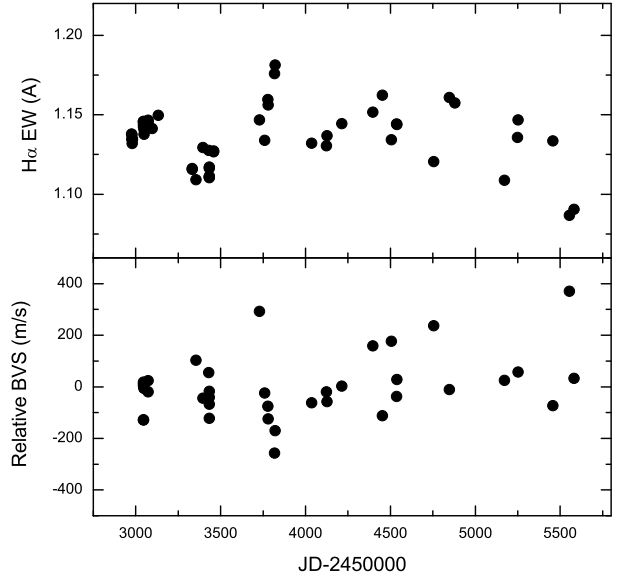}
      \caption{The examinations of the origin of the RV variations for HD 66141.
      JD vs. H$_{\alpha}$ EW variations (\emph{top panel}) and the BVS variations (\emph{bottom panel}), respectively, from December 2003 to January 2011.
        }
        \label{origin1}
   \end{figure}

\begin{figure}
   \centering
   \includegraphics[width=8cm]{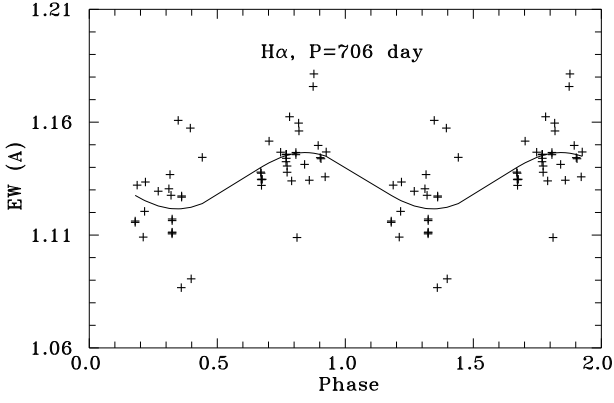}
      \caption{H$_{\alpha}$ EW variations for HD~66141 phased to the period of 706 days. The same epoch of JD 2448098.0 found in the 1.25$\mu$ (703-day flux variations) was used.
        }
        \label{Ha-phase}
   \end{figure}

Because the H$_{\alpha}$ absorption originates in the upper layers of stellar atmosphere and is also sensitive to stellar activity (K{\"u}rster et al. 2003), we used the H$_{\alpha}$ EW to measure the variations. Owing to sparse blending lines, a weak telluric line, and a narrow H$_{\alpha}$ absorption line (Figure~\ref{Ha1}), it is easy to estimate the H$_{\alpha}$ EW.
We measured the EW using a band pass of $\pm$ 1.0 ${\AA}$ centered on the core of the H$_{\alpha}$ line to avoid nearby blending lines (i.e. Ti\,I 6561.3, Na\,II 6563.9, and ATM H$_{2}$O 6564.0 ${\AA}$).
The mean EW of the H$_{\alpha}$ line in HD 66141 is measured to be 1135.7 $\pm$ 18.5 m${\AA}$. The rms of 18.5 m${\AA}$ corresponds to a 1.6\% variations in the EW. Figure~\ref{origin1} (top panel) shows the H$_{\alpha}$ EW variations as a function of time and the Lomb-Scargle periodogram of the H$_{\alpha}$ EW variations is shown in the middle panel of Figure~\ref{origin2}. There is large power at the frequency of 0.001416 $\pm$ 0.000070 c d$^{-1}$ with an FAP of less than 10$^{-5}$, corresponding to a period of 706.4 $\pm$ 35.0 days. Figure\,\ref{Ha-phase} shows the  H$_{\alpha}$ EW variations phased to a period of 706 days.

%
%
%

%
%

%
   \begin{figure}
   \centering
   \includegraphics[width=8cm]{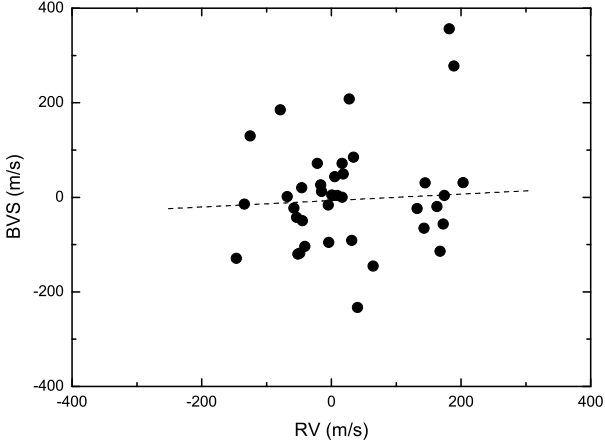}
      \caption{BVS vs. RV variations for HD 66141 from December 2003 to January 2011. The dashed line marks a slope of 0.07.
        }
        \label{BVS}
   \end{figure}
\subsection{Line bisector variations}

The RV variations produced by the rotational modulation of the surface inhomogeneities should produce some changes in the spectral line shape, such as line asymmetry (Queloz et al. 2001). Thus, the variations in the shapes of spectral lines help to interpret the origin of RV variations. The difference in the bisectors of line widths between the top and bottom of the line profile is defined as the bisector velocity span (BVS).

\begin{figure}
   \centering
   \includegraphics[width=8cm]{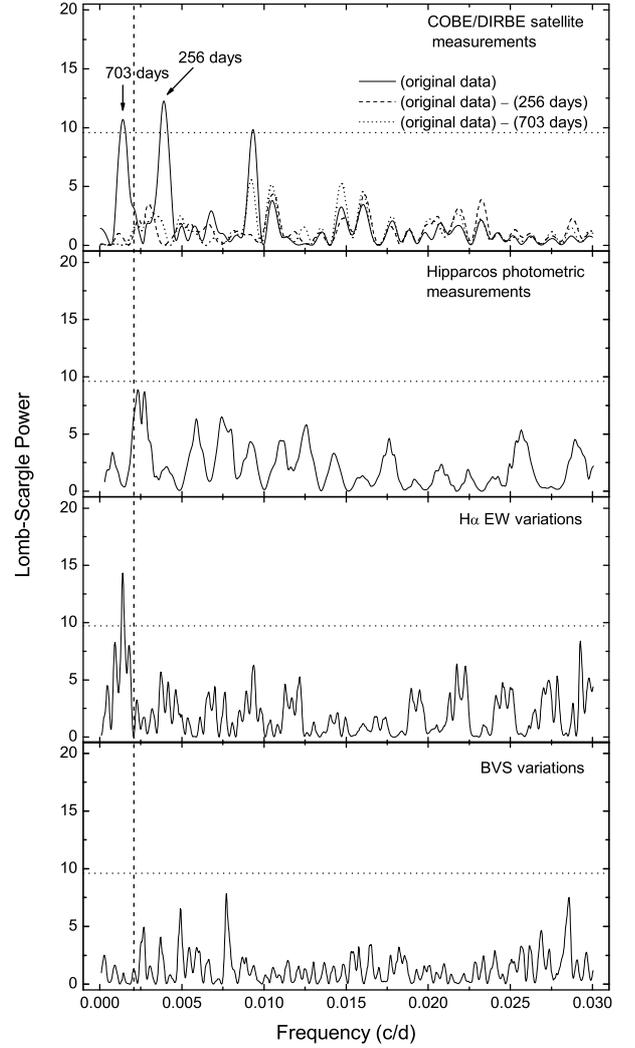}
      \caption{The Lomb-Scargle periodograms of the COBE/DIRBE satellite 1.25$\mu$ flux intensity measurements, the Hipparcos photometric measurements, the H$_{\alpha}$ EW variations, and the BVS variations for HD 166141 (\emph{top} to \emph{bottom panel}), respectively.
      The vertical dashed line marks the location of the period of 480 days and the horizontal dotted lines indicate an FAP threshold of 1 $\times 10^{-3}$ (0.1\%).
      \emph{Top panel} -- The solid line is the Lomb-Scargle periodogram of the COBE/DIRBE satellite 1.25$\mu$ flux intensity measurements for 3.5 years. The arrows show the positions of the dominant peaks at $f_{\rm 1}^{\rm IR}$ = 0.00390 $\pm$ 0.00008 c\,d$^{-1}$ ($P_{\rm 1}^{\rm IR}$ = 256.1 $\pm$ 5.4 days) and at $f_{\rm 2}^{\rm IR}$ = 0.00142 $\pm$ 0.00008 c\,d$^{-1}$ ($P_{\rm 2}^{\rm IR}$ = 703.0 $\pm$ 39.4 days). The dashed line shows the periodogram of the residuals after removal $P_{\rm 1}^{\rm IR}$ = 256 days fit to the original data. The dotted line shows the periodogram after removal $P_{\rm 2}^{\rm IR}$ = 703 days fit to the original data.
        }
        \label{origin2}
   \end{figure}

We measured the BVS using the least squares deconvolution (LSD) technique (Donati et al. 1997; Reiners \& Royer 2004; Glazunova et al. 2008), which is calculated by the mean profile of the spectral lines.
We also used the Vienna Atomic Line Database (VALD; Piskunov et al. 1995) to prepare the list of spectral lines. A total of $\sim$ 3900 lines within the wavelength region of 4500 -- 4900 {\AA} and 6450 -- 6840 {\AA} were used to construct the LSD profile, which excluded spectral regions around the I$_{2}$ absorption region, hydrogen lines, and regions with strong contamination by terrestrial atmospheric lines. Then, we calculated the BVS of the mean profile between two different central depth levels, 0.8 and 0.25. BVS variations as a function of time are exhibited in Figure~\ref{origin1} (bottom panel) and BVS vs. RV variations are exhibited in Figure~\ref{BVS}. We measured a slope of 0.07, which shows no correlation between BVS and the measured RV. The short-term accuracy of BVS measurements was measured by comparison of several spectra obtained in the same or during consecutive nights and is value of 45 m s$^{-1}$. The long-term BVS scatter might be indeed larger than that of the short-term measurements. We found a scatter of 116 m s$^{-1}$ for the whole BVS measurements of HD 66141. In Figure~\ref{origin2} (bottom panel), the Lomb-Scargle periodogram of the BVS shows no statistically significant peaks in the domain of interest.

%
\subsection{Orbital fit}

%
   \begin{figure}
   \centering
   \includegraphics[width=8cm]{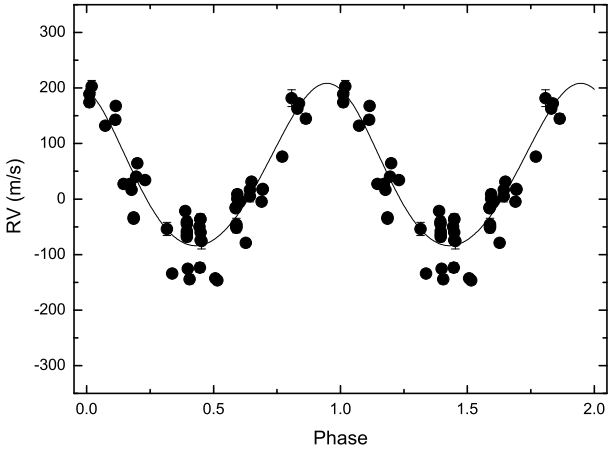}
      \caption{RV measurements for HD 66141 phased to the orbital period of 480.5 days.
      The solid line is the orbital solution that fits the data with a rms of 38.8 m s$^{-1}$.
              }
         \label{phase}
   \end{figure}

We find the variation fitted best with a Keplerian orbit of a period $P$ = 480.5 $\pm$ 0.5 days, a semi-amplitude $K$ = 146.2 $\pm$ 2.7 m s$^{-1}$ and an eccentricity $e$ = 0.07 $\pm$ 0.03. Solid line in Figure~\ref{orbit} shows the RV curve as a function of time for HD 66141 and the residuals after extracting the main frequency. The RV measurements phased to $P$ = 480.5 days are shown in Figure~\ref{phase}, and they exhibit clear periodic variations at this frequency. This Keplerian orbit determines the minimum mass of a planetary companion $m$ sin $i$ = 6.0 $\pm$ 0.3 $M_{\rm Jup}$ at a distance $a$ = 1.2 $\pm$ 0.1 AU from HD 66141. All the orbital elements are listed in Table~\ref{tab3}.

The dispersion of the RV residuals is 38.8 m s$^{-1}$, which is significantly larger than the rms scatter of the RV standard star (6.7 m s$^{-1}$) and the typical RV measurements error ($\sim$ 8 m s$^{-1}$).
We thus can propose that the large scatter in the residuals can be attributed to unresolved oscillations. Using the scaling relation (equation [7] from Kjeldsen \& Bedding 1995) with a luminosity and a mass of HD66141 (Table~\ref{tab2}), we expect the semi-amplitude of pulsations to be 39 m s$^{-1}$, which is almost in agreement with 38.8 m s$^{-1}$ dispersion of the residuals.

The visual inspection of the residuals shows also a long-period about $\pm$ 50 m s$^{-1}$ variations with a time-scale comparable to the time span of observations. It is premature to discuss the origin of long-term variations until we get longer time-span observations covering several variability cycles. For completeness, we would like to mention two other exoplanet-hosting candidate K giant stars from our survey: $\gamma^1$\,Leo (Han et al. 2010) and $\alpha$\,Ari (Kim et al. 2006; Lee et al. 2011). They also exhibit long-term variations of the RV residuals that are superimposed with short-term pulsational variability well established from night-to-night observations.

%
\begin{table}
\begin{center}
\caption{Orbital parameters of the best fit Keplerian orbit for HD 66141 b.}
\label{tab3}
\begin{tabular}{lc}
\hline
\hline
    Parameter                            & Value                             \\

\hline
    Period (days)                        & 480.5  $\pm$ 0.5                  \\
    $\it T$$_{\rm periastron}$ (JD)      & 2451320.8 $\pm$ 4.8 (1999.386)    \\
    $\it{K}$ (m s$^{-1}$)                & 146.2  $\pm$ 2.7                  \\
    $\it{e}$                             & 0.07   $\pm$ 0.03                 \\
    $\omega$ (deg)                       & 22.1   $\pm$ 3.5                  \\
    $f(m)$ ($\it M_{\odot}$)             & (1.545) $\times$ 10$^{-7}$        \\
    $a$ sin $i$ (AU)                     & (6.444) $\times$ 10$^{-3}$        \\
    $\sigma$ (O-C) (m s$^{-1}$)          & 38.8                              \\
\hline
    with $\textit{$M_{\star}$}$ = 1.1 $\pm$ 0.1 ($M_{\odot}$)      &         \\
    $m$ sin $i$ ($\it M_{\rm Jup}$)      & 6.0 $\pm$ 0.3                     \\
    $\it{a}$ (AU)                        & 1.2 $\pm$ 0.1                     \\
\hline

\end{tabular}
\end{center}
\end{table}

%

\section{Discussion and conclusion}

From the analysis of the eight-year precise RV measurements, we found compelling evidence for a low-amplitude and long-period 480-day RV variations in a K giant HD 66141. We examined possible origins of these RV variations. The Ca II H line profile, the Hipparcos photometric measurements, and the BVS measurements show no detectable indication of variability at this period.

However, the analysis of 1.25$\mu$ COBE/DIRBE infrared flux measurements for HD~66141 revealed two periods of 256 or 703 days that well fit the observed flux variations. Judging which a period is real came from the EW analysis of H$_{\alpha}$ line, which revealed almost the same period of variations (703.0 $\pm$ 39.4 vs. 706.4 $\pm$ 35.0 days). Comparison of Figure\,\ref{cobe} and Figure\,\ref{Ha-phase} clearly shows about 0.18 phase lag between the time of maximum H$_{\alpha}$ EW and the maximum of the 1.25$\mu$ flux.
We note that the wavelength-to-wavelength phase lags are clearly present in variations of M giants and Mira variables with the optical maximum appearing about 0.18 phase before the maxima at 1.25 $\mu$ (Price et al. 2010). It was shown by Alvarez \& Plez (1998) that optical/NIR lags in oxygen-rich Miras are likely due to  strong titanium oxide (TiO) variability during a pulsation cycle. For the case of K2 III star HD 66141 the TiO lines might be formed in a cool spot on the surface having the temperature contrast of about 1000 K. Thus, the temperature in the cool spot might approach to about 3500 K (the typical temperature of M giants). The question on the origin of 0.18 phase lag between H$_{\alpha}$ EW and the 1.25$\mu$ flux variations need further confirmations and investigations.

Returning to about 703 -- 706 day periods of variations in the NIR fluxes and optical absorptions in a spectral line we note that it stands far aside the RV period of 480 days. The most likely origin of 703 -- 706 day periods is certainly the surface spot(s) and the rotation of HD 66141. These periods are equal to each other within accuracy, thus we deduce a mean value $P_{\mathrm{rot}}$ = 705 $\pm$ 53 days as the rotation period of star. With a new estimation of rotation period of HD 66141 and accurate radius, we can accurately re-estimated  the rotational velocity as $v_{\rm rot}$ sin $i$ = 1.5 km~s$^{-1}$.

In summary, the most likely cause of 480-day RV variations is a planet orbiting the star. HD 66141 is a planetary system with a 1.1 $\pm$ 0.1 $\it M_{\odot}$ giant star and a 6.0 $\pm$ 0.3 $\it M_{\rm Jup}$ planet.




\begin{acknowledgements}
     BCL acknowledges partial support by the KASI (Korea Astronomy and Space Science Institute) grant 2012-1-410-03. Support for MGP was provided by the National Research Foundation of Korea to the Center for Galaxy Evolution Research. DEM acknowledges his work as part of the research activity of the National Astronomical Research Institute of Thailand (NARIT), which is supported by the Ministry of Science and Technology of Thailand. We thank the developers of the Bohyunsan Observatory Echelle Spectrograph (BOES) and all staff of the Bohyunsan Optical Astronomy Observatory (BOAO). We are grateful to the anonymous referee for useful comments that have greatly improved the quality of the manuscript. This research made use of the SIMBAD database, operated at the CDS, Strasbourg, France.
\end{acknowledgements}
%



\begin{thebibliography}{}

\bibitem[Alvarez \& Plez(1998)]{1998A&A...330.1109A} Alvarez, R., \& Plez, B.\ 1998, \aap, 330, 1109

\bibitem[Batten(1983)]{1983BICDS..24....3B} Batten, A.~H.\ 1983, Bulletin Information du Centre de Donnees Stellaires, 24, 3

\bibitem[Butler et al.(1996)]{1996PASP..108..500B} Butler, R.~P., Marcy, G.~W., Williams, E., et al.\ 1996, \pasp, 108, 500

\bibitem[da Silva et al.(2006)]{2006A&A...458..609D} da Silva, L., Girardi, L., Pasquini, L., et al.\ 2006, \aap, 458, 609

\bibitem[de Medeiros \& Mayor(1999)]{1999A&AS..139..433D} de Medeiros, J.~R., \& Mayor, M.\ 1999, \aaps, 139, 433

\bibitem[D{\"o}llinger et al.(2009)]{2009A&A...505.1311D} D{\"o}llinger, M.~P., Hatzes, A.~P., Pasquini, L., et al.\ 2009, \aap, 505, 1311

\bibitem[Donati et al.(1997)]{1997MNRAS.291..658D} Donati, J.-F., Semel, M., Carter, B.~D., et al.\ 1997, \mnras, 291, 658

\bibitem[Eaton \& Williamson(2007)]{2007PASP..119..886E} Eaton, J.~A., \& Williamson, M.~H.\ 2007, \pasp, 119, 886

\bibitem[Endl et al.(2000)]{2000A&A...362..585E} Endl, M., K{\"u}rster, M., \& Els, S.\ 2000, \aap, 362, 585

\bibitem[ESA(1997)]{1997yCat.1239....0E} ESA 1997, VizieR Online Data Catalog, 1239, 0

\bibitem[Famaey et al.(2005)]{2005A&A...430..165F} Famaey, B., Jorissen, A., Luri, X., et al.\ 2005, \aap, 430, 165

\bibitem[Fekel(1997)]{1997PASP..109..514F} Fekel, F.~C.\ 1997, \pasp, 109, 514

\bibitem[Frink et al.(2002)]{2002ApJ...576..478F} Frink, S., Mitchell, D.~S., Quirrenbach, A., et al.\ 2002, \apj, 576, 478

\bibitem[Galazutdinov (1992)]{} Galazutdinov, G. A. 1992, Special Astrophysical Observatory Preprint 92 (Nizhnij Arkhyz: SAO)

\bibitem[Girardi et al.(2000)]{2000A&AS..141..371G} Girardi, L., Bressan, A., Bertelli, G., et al.\ 2000, \aaps, 141, 371

\bibitem[Glazunova et al.(2008)]{2008AJ....136.1736G} Glazunova, L.~V., Yushchenko, A.~V., Tsymbal, V.~V., et al. \ 2008, \aj, 136, 1736

\bibitem[Han et al.(2007)]{2007PKAS...22...75H} Han, I., Kim, K.-M., Lee, B.-C., et al.\ 2007, PKAS, 22, 75

\bibitem[Han et al.(2008)]{2008JKAS...41...59H} Han, I., Lee, B.-C., Kim, K.-M., et al.\ 2008, JKAS, 41, 59

\bibitem[Han et al.(2010)]{2010A&A...509A..24H} Han, I., Lee, B.~C., Kim, K.~M., et al.\ 2010, \aap, 509, A24

\bibitem[Hatzes \& Cochran(1993)]{1993ApJ...413..339H} Hatzes, A.~P., \& Cochran, W.~D.\ 1993, \apj, 413, 339

\bibitem[Hatzes \& Cochran(1998)]{1998ASPC..154..311H} Hatzes, A.~P., \& Cochran, W.~D.\ 1998, ASPC, 154, 311

\bibitem[Hatzes et al.(2005)]{2005A&A...437..743H} Hatzes, A.~P., Guenther, E.~W., Endl, M., et al.\ 2005, \aap, 437, 743

\bibitem[Hatzes et al.(2006)]{2006A&A...457..335H} Hatzes, A.~P., Cochran, W.~D., Endl, M., et al.\ 2006, \aap, 457, 335

\bibitem[Johnson et al.(2007)]{2007ApJ...665..785J} Johnson, J.~A., Fischer, D.~A., Marcy, G.~W., et al.\ 2007, \apj, 665, 785

\bibitem[J{\o}rgensen \& Lindegren(2005)]{2005A&A...436..127J} J{\o}rgensen, B.~R., \& Lindegren, L.\ 2005, \aap, 436, 127

\bibitem[Kharchenko et al.(2007)]{2007AN....328..889K} Kharchenko, N.~V., Scholz, R.-D., Piskunov, A.~E., et al. \ 2007, Astronomische Nachrichten, 328, 889

\bibitem[Kim et al.(2006)]{2006A&A...454..839K} Kim, K.~M., Mkrtichian, D.~E., Lee, B.-C., et al.\ 2006, \aap, 454, 839

\bibitem[Kim et al.(2007)]{2007PASP..119.1052K} Kim, K.-M., Han, I., Valyavin, G.~G., et al.\ 2007, \pasp, 119, 1052

\bibitem[Kjeldsen \& Bedding(1995)]{1995A&A...293...87K} Kjeldsen, H., \& Bedding, T.~R.\ 1995, \aap, 293, 87

\bibitem[K{\"u}rster et al.(1999)]{1999A&A...344L...5K} K{\"u}rster, M., Hatzes, A.~P., Cochran, W.~D., et al.\ 1999, \aap, 344, L5

\bibitem[K{\"u}rster et al.(2003)]{2003A&A...403.1077K} K{\"u}rster, M., Endl, M., Rouesnel, F., et al.\ 2003, \aap, 403, 1077

\bibitem[Lee et al.(2008)]{2008AJ....135.2240L} Lee, B.-C., Mkrtichian, D.~E., Han, I., et al.\ 2008, \aj, 135, 2240

\bibitem[Lee et al.(2011)]{2011A&A...529A.134L} Lee, B.-C., Mkrtichian, D.~E., Han, I., et al.\ 2011, \aap, 529, A134

\bibitem[Lomb(1976)]{1976Ap&SS..39..447L} Lomb, N.~R.\ 1976, \apss, 39, 447

\bibitem[Lovis \& Mayor(2007)]{2007A&A...472..657L} Lovis, C., \& Mayor, M.\ 2007, \aap, 472, 657

\bibitem[Massarotti et al.(2008)]{2008AJ....135..209M} Massarotti, A., Latham, D.~W., Stefanik, R.~P., et al.\ 2008, \aj, 135, 209

\bibitem[Niedzielski et al.(2007)]{2007ApJ...669.1354N} Niedzielski, A., Konacki, M., Wolszczan, A., et al.\ 2007, \apj, 669, 1354

\bibitem[Pasquini et al.(1988)]{1988A&A...191..253P} Pasquini, L., Pallavicini, R., \& Pakull, M.\ 1988, \aap, 191, 253

\bibitem[Pearce (1955)]{1955Trans.IAU,..IX..442} Pearce, J.A. \ 1955, Trans. IAU. XI, 442

\bibitem[Perryman et  al.(1997)]{1997A&A...323L..49P} Perryman, M.~A.~C., Lindegren, L., Kovalevsky, J., et al.\ 1997, \aap, 323, L49

\bibitem[Piskunov et al.(1995)]{1995A&AS..112..525P} Piskunov, N.~E., Kupka, F., Ryabchikova, T.~A., et al. \ 1995, \aaps, 112, 525

\bibitem[Price et al.(2010)]{2010ApJ..190..203P} Price, S.~D., Smith,B.~J., Kuchar, T.~A., et al. \ 2010, \apj, 190, 203

\bibitem[Queloz et al.(2001)]{2001A&A...379..279Q} Queloz, D., Henry, G.~W., Sivan, J.~P., et al.\ 2001, \aap, 379, 279

\bibitem[Reffert et al.(2006)]{2006ApJ...652..661R} Reffert, S., Quirrenbach, A., Mitchell, D.~S., et al.\ 2006, \apj, 652, 661

\bibitem[Reiners \& Royer(2004)]{2004A&A...415..325R} Reiners, A., \& Royer, F.\ 2004, \aap, 415, 325

\bibitem[Richichi \& Percheron(2002)]{2002A&A...386..492R} Richichi, A., \& Percheron, I.\ 2002, \aap, 386, 492

\bibitem[Saar \& Donahue(1997)]{1997ApJ...485..319S} Saar, S.~H., \& Donahue, R.~A.\ 1997, \apj, 485, 319

\bibitem[Sato et al.(2003)]{2003ApJ...597L.157S} Sato, B., Ando, H., Kambe, E., et al.\ 2003, \apjl, 597, L157

\bibitem[Scargle(1982)]{1982ApJ...263..835S} Scargle, J.~D.\ 1982, \apj, 263, 835

\bibitem[Setiawan et al.(2003)]{2003A&A...398L..19S} Setiawan, J., Hatzes, A.~P., von der L{\"u}he, O., et al.\ 2003, \aap, 398, L19

\bibitem[Takeda et al.(2005)]{2005PASJ...57...27T} Takeda, Y., Ohkubo, M., Sato, B., et al.\ 2005, \pasj, 57, 27

\bibitem[Tody(1986)]{1986SPIE..627..733T} Tody, D.\ 1986, \procspie, 627, 733

\bibitem[Udry et al. (1999)]{1999ASPC..185..383U}Udry S., et al., 1999, ASPC, 185, 383

\bibitem[Udry, Mayor, \& Queloz(1999)]{1999ASPC..185..367U} Udry S., Mayor M., Queloz D., 1999, ASPC, 185, 367

\bibitem[van Leeuwen(2007)]{2007A&A...474..653V} van Leeuwen, F.\ 2007, \aap, 474, 653


\end{thebibliography}
\end{document}